\documentclass[10pt,conference]{IEEEtran}
 \pdfoutput=1
\IEEEoverridecommandlockouts
\usepackage{booktabs} 
\usepackage{diagbox}
\usepackage{adjustbox}
\usepackage{hyperref}
\usepackage{hhline}
\usepackage{multirow}
\usepackage{flushend}

\newenvironment{boxedtext}
    {
    
    \begin{center}

    \begin{tabular}{|p{0.96\linewidth}|}
    \hline
    }
    { 
    \\ \hline
    \end{tabular} 
    
    \end{center}
    \vspace{-5pt}
       }

\newcolumntype{C}[1]{>{\centering\let\newline\\\arraybackslash\hspace{0pt}}m{#1}}
\newcolumntype{R}[1]{>{\hfill\let\newline\\\arraybackslash\hspace{0pt}}m{#1}}
\newcommand{\graycell}[1]{ {\textbf{#1}}}

\makeatletter
\edef\orig@output{\the\output}
\output{\setbox\@cclv\vbox{\unvbox\@cclv\vspace{0pt plus 20pt}}\orig@output}
\makeatother

\begin{document}

\title{Expressions of Sentiments During Code Reviews: Male vs. Female}

\author{
       \IEEEauthorblockN{
           Rajshakhar Paul%
           \IEEEauthorrefmark{1},
           Amiangshu Bosu%
           \IEEEauthorrefmark{1},
           Kazi Zakia Sultana%
           \IEEEauthorrefmark{2},
            }
       \IEEEauthorblockA{
           \IEEEauthorrefmark{1}
           Department of Computer Science\\
           Wayne State University,
           Detroit, MI, USA\\
          }           
       \IEEEauthorblockA{%
       \IEEEauthorrefmark{2}
       Department of Computer Science and Engineering\\
           Montclair State University,
           Montclair, NJ, USA\\
          Email:~
           r.paul@wayne.edu,
           amiangshu.bosu@wayne.edu,
           sultanak@montclair.edu}
   }   
\maketitle

\begin{abstract}
\textit{Background:} 
As most of the software development organizations are male-dominated, female developers encountering various negative workplace experiences reported feeling like they ``do not belong''. Exposures to discriminatory expletives  or negative critiques from their male colleagues may further exacerbate those feelings.

\textit{Aims:}  The primary goal of this study is \textit{to identify the differences in expressions of sentiments between male and female developers during various software engineering tasks.} 

\textit{Method:} On this goal, we mined the code review repositories of six popular open source projects. We used a semi-automated approach leveraging the name as well as  multiple social networks to identify the gender of a developer. Using SentiSE, a customized and state-of-the-art sentiment analysis tool for the software engineering domain, we classify each communication as negative, positive, or neutral. We also compute the frequencies of sentiment words, emoticons, and expletives used by each developer.

\textit{Results:} Our results suggest that the likelihood of using sentiment words, emoticons, and expletives during code reviews varies based on the  gender of a developer, as  females are significantly less likely to  express sentiments than males. Although female developers were more neutral to their male colleagues than to another female, male developers from three out of the six projects  were not only writing more frequent negative comments but also withholding positive encouragements from their female counterparts.

\textit{Conclusion:} Our results provide empirical evidence of another factor behind the negative work place experiences encountered by the female developers  that may be contributing to the diminishing number of females in the SE industry.

\end{abstract}


\section{Introduction}
\label{sec-introduction}
Until 1960's, programming was considered women's job and the computing area was dominated by women~\cite{abbate2012recoding}. However, as the demand for programmers increased,  male programmers sought to create their dominance through creating professional associations, through ad campaigns discouraging the hiring of women, and by adding personality tests biased against female applicants~\cite{ensmenger2012computer}.
As the area of computing has thrived over the last five decades, the participation of women in computing has declined.  According to the US Bureau of Labor Statistics, females accounted for about 37\% of the U.S. college students who received bachelor's degrees in Computer and Information Sciences (CIS) in 1986~\cite{ensmenger2012computer}; twenty years later, in 2006, that number declined to only 12\%~\cite{taulbee2006}. The percentage of females in the software industry also declined rapidly during the same period~\cite{platman2004workforce} and most of the contemporary software development organizations are male-dominated~\cite{panteli1999status}.

Recently, female software developers reported not only sexual harassment~\cite{das2009sexual} but also various technical biases. For example, women are often assigned menial tasks~\cite{didio1997} and code commits from females are less likely to be accepted~\cite{terrell2017gender} than males.  Marwick \textit{et} al. claims that in the current software industry, females often feel like they ``\textit{do not belong}''  and often struggle with an ``\textit{imposter syndrome}''\footnote{ A psychological phenomenon in which they feel a sense of inadequacy despite being perfectly competent.}~\cite{marwick2017silicon}. 
Exposures to discriminatory expletives  or negative critiques from their male colleagues may further exacerbate those feelings.

There are several anecdotal evidence suggesting demeaning attitude from males towards female developers. For example, a female developer from the Github mentioned in an interview with the Techcrunch~\cite{techcrunch},``\textit{.. really hard time getting used to the culture, the aggressive communication on pull requests and how little the men I worked with respected and valued my opinion.}" She ``\textit{participated in the boys’ club upon joining,}'' but later found ``\textit{her character being discussed in inappropriate places like on pull requests and issues.}''
Another memo~\cite{google-memo} written by a Google employee claimed ``\textit{Differences in distributions of traits between men and women may in part explain why we don't have 50 percent representation of women in tech and leadership.}''
However, no study has yet investigated whether or not, female developers are more frequent recipients of negative opinions than their male colleagues. We believe such an investigation can possibly determine another factor behind the low participation of female developers in the contemporary software industry. Therefore, the primary objective of this study is \textit{to identify the differences in expressions of sentiments between male and female developers during various software engineering tasks.} 
However, the goal of our research is not to investigate the differences between the inherent writing styles of males and females, rather we aim to find out if they express sentiments differently during  software engineering interactions and whether their opinions differ based on the gender of the recipient. If we find differences, in a followup study, we want to investigate whether the behavior is resulting from any social context or gender bias or from a technological standpoint.

Expressions of positive or negative opinions during various Software Engineering (SE) activities are common as prior SE research found sentiments in commit messages~\cite{guzman2014sentiment}, issue tracking systems~\cite{murgia2014developers}, code review comments~\cite{ahmed2017senticr}, Stack Overflow posts~\cite{calefato2017sentiment} and  mailing-lists~\cite{Guzman2013}. 
While no study has yet explored the differences in sentiment expressions between male and female software developers, research in other domains suggest existences of significant differences based on the gender of an author. For example, females express significantly more positive sentiments on social media~\cite{thelwall2010data}, use emoticons  differently than males~\cite{wolf2000emotional}, and emotion-based features in micro-blog posts can be used to predict the gender of an author~\cite{montero2014investigating}.
As prior research shed lights on the differences of opinions between males and females in other domains, we are motivated to investigate this question in the SE field.

Among the various software engineering interactions, such as code reviews, bug discussions, code commits, and StackOverflow discussions, this study focuses on code reviews. 
We selected code reviews as those facilitate direct communications between developers and are one of the primary interactions where developers may express sentiments~\cite{Bosu-TSE-full}. For this research, we mined the code review repositories of six popular open source software (OSS) projects and identified all the developers that have committed at least five code changes for those projects. Based on the approach adopted in a recent study~\cite{terrell2017gender}, we developed a semi-automated methodology followed by a manual validation using social networks (i.e., LinkedIn, Google Plus, Facebook, Github, and Twitter) to identify the genders of the `Non-casual developer'\footnote{A developer who has submitted at least five code changes to a project under our study.}.  We applied SentiSE, a customized and state-of-the-art sentiment analysis tool for the SE domain to identify the sentiment polarity of each code review comment.  
  
In summary, the primary contributions of this study are:
\begin{itemize}
  \item A comparison of the opinions expressed by male and female software developers during code reviews.
    \item A comparison of the  different categories of sentiment words used by male and female developers. 
  \item A comparison of the emoticons, expletives and swear words used by male and female developers.  
  \item A comparison of the opinions expressed during  same-gender and cross-gender interactions.
  \item An empirical evidence of the differences between male and female developers during SE interactions.
\end{itemize}

The remainder of the paper is organized as follows. 
Section \ref{sec-background} provides background about code review and sentiment analysis.
Section \ref{sec-research-question} introduces the research questions of this study. 
Section \ref{sec-research-methodology} describes our  research methodology. 
Section \ref{sec-results} presents the results of this study. 
Section \ref{sec-discussion} discusses the implications of the results. 
Section \ref{sec-threats} describes the threats to validity of our findings. 
Finally, Section \ref{sec-conclusion} provides some directions for future work and concludes the paper.

\section{Background}
\label{sec-background}
This section presents a brief background on three topics relevant to this study: Gender issues in Software Engineering, Sentiment analysis, and peer code reviews.

\subsection{Gender Issues in Software Engineering}
Earlier studies in software engineering focused on the participation of women in Free/Libre/ Open Source Software (FLOSS) projects and reported extremely lower (i.e., between 2\% to 5\%) ratio of female contributors~\cite{ghosh2002free}. The reason behind the lower participation of female is attributed to the social and cultural arrangements of FLOSS projects, which actively excludes female contributors~\cite{nafus2006gender}. 
Due to several initiatives from few of the FLOSS projects (i.e., Debian, and GNOME) to attract female contributors, the number of female contributors has improved to around 11\% in recent years~\cite{Robles2016}; however the number is still less than half of the industry average (i.e., 24\%). 

Similar to other domains \cite{campbell2008gender,herring2009does}, SE researchers have also observed the positive impacts of gender diversity on software development team productivity~\cite{vasilescu2015gender, ortu4diverse}. 
Yet, studies also found various types of discrimination against women. For example, women in computing organizations are  often assigned menial tasks, while similarly male colleagues are given `choice' projects ~\cite{didio1997}, often do not get opportunities at management positions,  and earn lower salaries compared to men~\cite{panteli1999status}. Women also often receive unfair evaluation as 72\% women reported sensing gender bias in how they were evaluated~\cite{hewlett2014athena}.   As a result, female developers become increasingly pessimistic about career opportunities as their tenure progress~\cite{james2017perception}. Even in FLOSS projects, where most of the participants are volunteers, women perform majority other types of contributions than coding, while men mostly contribute with code~\cite{Robles2016}. A recent study on Github reports that although women significantly have higher acceptance rate then men in general, if a woman's gender is identifiable from her profile, her pull requests are less likely to be accepted~\cite{terrell2017gender}.

\subsection{Sentiment Analysis}
\textit{Sentiment analysis} is the computational process of determining the emotional tone behind a series of words to gain an understanding of the attitudes, opinions and emotions of a speaker or an author~\cite{liu2012survey}.
Sentiment analysis applications have spread to almost every possible domain, from consumer products, services, healthcare, and financial services to social events and political elections~\cite{feldman2013techniques}.

SE researchers have recently applied sentiment analysis on various SE artifacts and found developers expressing sentiments in code commits, issue reports, and project forums. 
Pleatea \textit{et al.}~\cite{pletea2014security} analyzed sentiments of  discussions on  GitHub using NLTK \cite{bird2006nltk} and found more expressions of negative sentiments in security-related discussions than in  other  discussions. 
Guzman \textit{et al.}~\cite{guzman2014sentiment} used SentiStrength to analyze commit messages on Github and observed projects developed in Java having more negative commit messages than projects developed in other languages. They also observed higher expressions of positive sentiments among distributed teams. 
Garcia et al.~\cite{garcia2013} carried out a similar analysis on both the issue reports and mailing-list discussions of the Gentoo community. They observed significant correlations between the activities of authors and their expressions of emotions.
Murgia \textit{et al.}~\cite{murgia2014developers} analyzed issue reports from the Apache Foundation projects and observed developers expressing sentiments during bug discussions. 

Most of the prior works focused on identifying sentiments expressed during various SE activities.
Couple of recent works have evaluated the impact of those sentiments on the outcomes of different activities.
Ortu \textit{et al.}~\cite{ortu2015bullies} analyzed the relation between bug resolution time with sentiment, emotions and politeness of developers and found that issues with negative sentiments are often associated with longer resolution times.
Islam \textit{et al.}~\cite{islam2016exploration} investigated emotional variations during various development activities (e.g., bug-fixing tasks) and found that emotional active developers tend to post longer commit messages.

\subsection{Code Review}
Peer code review is a software engineering practice, where a developer sends his/her code to a peer to  identify possible defects before merging to the project codebase.
Compared with the traditional heavy-weight inspection process, peer code review is more informal, tool-based, and used regularly in practice~\cite{bacchelli2013}.  
To make peer code reviews more efficient, teams use automated support tools such as  Gerrit\footnote{\url{https://code.google.com/p/gerrit/}}, Phabricator\footnote{\url{http://phabricator.org/}}, and ReviewBoard\footnote{\url{https://www.reviewboard.org/}}. 
A tool-based code review process starts when an author creates a \emph{patchset} (i.e. all files added or modified in a single revision) along with a description of the changes and submits that information to a code review tool.
To facilitate reviews, code review tools highlight the changes between two revisions in a side-by-side display. 
Both the reviewers and the author can insert comments pointing out issues, suggesting improvements, or clarifying the change.
After the review, the author may upload a new patch-set addressing the review comments and initiate a new review iteration. 
This review cycle repeats until either the reviewers approve the change or the author abandons it.
Code review tools capture the interactions (a.k.a. review comments) between the author and a reviewer to facilitate \textit{post-hoc} analyses.

\section{Research Questions}
\label{sec-research-question}
The primary objective of this study is \textit{to identify the differences in expressions of sentiments between male and female developers during various software engineering tasks.}  Following subsections introduce five specific research questions to investigate this objective. As this study focuses specifically on code reviews, our research questions are aimed towards code review comments.

\subsection{Gender Vs. Sentiment}

Traditional believes in the  Western culture portrait women as ``the emotional sex'' and consider men as emotionally inexpressive compared to women~\cite{zammuner2000men}. Recent research on social media posts~\cite{montero2014investigating,thelwall2010data} also found females expressing more sentiments than males. Therefore, our first research question investigates if these findings are also applicable to professional workplace interactions such as code reviews.

\begin{em}
\begin{description}
\item [RQ1:] Does the likelihood of expressing sentiments during code reviews depend on the gender of a developer?
\end{description}
\end{em}

\subsection{Same Gender Vs. Cross-Gender Interaction}

Prior works found significant differences between same-gender and cross-gender interactions~\cite{wolf2000emotional,stapleton2003gender}. In general, we would expect a developer to be less expressive, when interacting with a person from the opposite gender than with a person from the same gender. Our next research question investigates this expectation. 

\begin{em}
\begin{description}
\item [RQ2:] Do developers express sentiments differently during their cross-gender interactions than during their same gender interactions?
\end{description}
\end{em}

\subsection{Gender Vs. Sentiment Word}



The study on gender and language began in the early 1970s and was arguably established by Robin Lakoff~\cite{lakoff1975}. The book worked on the foundational ideas about gendered language. She outlined some specific tendencies  of wording or writing styles in women's language. Although there are many controversies and criticisms against her work, this groundbreaking research unmasked the nature of male supremacy. Another recent work found females authoring more happy / sad tweets than males, while males' tweets expressed more surprises / fears~\cite{Volkova2016}. These works motivate us to investigate similar differences in the SE domain. Our next research question investigates the use of various categories of sentiment words (Table~\ref{table:sentimetExamples}) based on the  gender of a developer. 

\begin{em}
\begin{description}
\item [RQ3:] Do the categories of sentiment words used during code reviews vary based on the gender of a developer?
\end{description}
\end{em}

\subsection{Gender Vs. Emoticon}

Emoticons are widely used in informal written communication and help the author expressing his/her feelings or mood. Recent studies suggest significant differences based on gender with females using emoticons more frequently than males~\cite{Volkova2016,wolf2000emotional}. Therefore, we are interested to find out if these results are supported in a SE context.

\begin{em}
\begin{description}
\item [RQ4:] Does the likelihood of using emoticons during code reviews depend on the gender of a developer?
\end{description}
\end{em}

\subsection{Gender vs. Expletive/ Swear Word}

Swearing, or the use of expletives, are perceived as intrinsically forceful or aggressive activities. Cultural stereotype as well as the association of expletives with `masculine identity'~\cite{klerk1991expletives} suggest a less likelihood of expletives from females than from males. However, recent studies on a contemporary culture~\cite{stapleton2003gender, ott} did not find any significant differences. Our next question  explores whether female developers follow a traditional stereotype or a contemporary culture~\cite{stapleton2003gender}.  

\begin{em}
\begin{description}
\item [RQ5:] Does the likelihood of using swear words /expletives during code reviews depend on the gender of a developer?
\end{description}
\end{em}




\section{Research Methodology}
\label{sec-research-methodology}
The accurate identification of  the gender of a contributor is not only essential but also  challenging for this research. In the following subsections, we describe our data collection, gender resolution strategy, and sentiment analysis methodology.

\begin{table*}
	\caption{Project Demographics}
	\label{table:projects}
	\resizebox{\textwidth}{!}{
\begin{tabular}{|l|l|l|l|R{1.6cm}|R{1.5cm}|R{1.5cm}|R{1.6cm}|R{1.5cm}|}
\hline
\graycell{Project} & 
\graycell{Domain} & 
\graycell{Technology} & 
\multicolumn{1}{C{1.5cm}|}{ \graycell{Using Gerrit since}} & 
\multicolumn{1}{C{1.6cm}|}{\graycell{Requests mined*}} & 
\multicolumn{1}{C{1.5cm}|}{\graycell{ Total devs. }} &
\multicolumn{1}{C{1.5cm}|}{\graycell{Non-casual devs.}} &
\multicolumn{1}{C{1.6cm}|}{\graycell{\% CR by non-casual devs}}&
\multicolumn{1}{C{1.5cm}|}{\graycell{\% Female}}\\
 
\hline
Android & {Mobile OS} &C, C++, Java & October, 2008 & 81,137 & 2,589  & 981 & 95.6\% & 6.19\%   \\ 
\hline 
Chromium OS & {Desktop OS} &C, C++ & March, 2011 & 153,523  & 1,511  & 1,019 & 99.2\% & 8.74\%   \\ 
\hline 
Couchbase & {NoSQL database} &C, C++ & May, 2010 & 64,799  & 247 & 165 &  99.7\% & 9.69\%  \\ 

\hline 
OmapZoom & Mobile Platform & C & February, 2009 & 35,973  & 604 & 439 & 95.9\% & 7.06\% \\ 

\hline 
OVirt & Virtualization & Java &October, 2011 & 73,523  & 345 &220 & 99.4\% & 9.54\%   \\ 

\hline 
Qt & UI framework & C, C++ & May, 2011 & 155,936   & 1,598 & 746 & 94.9\% & 3.12\%   \\ 
\hline

\multicolumn{3}{l}{~~~*Mined during September, 2017} & 
\multicolumn{1}{|C{1.3cm}|}{\graycell{Total:}}&
\multicolumn{1}{R{1.5cm}|}{\graycell{564,891}}&
\multicolumn{1}{R{.8cm}|}{}&
\multicolumn{1}{R{1.5cm}|}{\graycell{3,570}}\\
\hhline{~~~--~-}
\end{tabular} 

}
\end{table*}

\subsection{Data Collection and Preparation}
We used the Gerrit-Miner tool of ~\cite{bosu2013impact} to mine completed code reviews of 12 popular OSS projects and stored the data in a MySQL database. 
Among the 12 projects, we excluded the six projects that did not satisfy either of the following two criteria: i) an open source project that mandates each and every change to be submitted for reviews on Gerrit; and ii) project contributors have performed at least 30,000 code reviews. The selected six projects had total 564,891 completed (i.e., `Merged' or `Abandoned') code reviews.
Table~\ref{table:projects} shows the list of projects. 

A manual inspection of the comments posted by some accounts (e.g., `Qt Sanity Bot' or `BuildBot') suggested that those accounts were automated bots rather than humans. These accounts typically contain one of the following keywords: `bot', `auto', `CI', `Jenkins', `integration', `build', `hook', `recheck', `travis', or `verifier'. Because we wanted only code reviews from actual reviewers, we excluded these bot accounts after a manual inspection had confirmed that the interactions were automatically generated.
Following a similar approach as Bird \textit{et al.}~\cite{bird2006mining}, we used the Levenshtein distance between two names to identify similar names. If our manual reviews of the associated accounts suggested that those belong to the same person we merged those to a single account.

In this study, we define a \textit{`Non-casual developer'} as a developer who has submitted at least five code changes for his/her project.
 Since our gender resolution  strategy is time consuming, we only considered the non-casual developers in each project for our subsequent analyses. 
 Column `Total devs.' and column `non-casual devs.' in Table~\ref{table:projects} show the total number of developers and the number of non-casual developers in each of the six projects respectively. Although, the number of non-casual developers may be as low as 38\%  of total developers (i.e., Android), they contributed more than 95\% of total code reviews in each of the six projects (column `\% CR by non-casual developers' in Table~\ref{table:projects}).
 Our data preparation steps generated a list of 3,570 non-casual developers (Table~\ref{table:projects}: Non-casual devs.) from the six projects.

\subsection{Gender Resolution}
We adopted a semi-automated gender resolution strategy using the  genderComputer tool created by Vasilescu \textit{et al.}~\cite{VasilescuIWC13} and modified by Terrell \textit{et al.}~\cite{terrell2017gender} and followed the automated steps with  manual validations using publicly available information on social networks. The genderComputer tool uses a database of 221,854 first names from 204 countries around the world and classifies each name belonging to one of the following four categories:
\begin{enumerate}
    \item \textit{Male}- names given to males.
    \item \textit{Female}- names given to females.
    \item \textit{Unisex} - names given to both males and females.
    \item \textit{None}- no entry in the database.
\end{enumerate}

Based on the names of our 3,570 non-casual developers from the six projects, the genderComputer tool classified  the contributors as following: 2,633 males, 325 females, 496 unisex, and 116 none. To ensure the accuracy of the identified genders, we adopted following five-step manual validation strategy for the 937 non-male (i.e., female, unisex, or none) contributors. We moved onto the next resolution step only if all the previous steps failed. If all the five manual validation steps were unsuccessful, which occurred for only 39 contributors $(\approx 1\%$), we excluded a developer from our subsequent analyses. Although, our gender resolution steps leverage only publicly available information, we got our research method reviewed and approved by our Institutional Review Board (IRB).

\subsubsection{Resolution using Gerrit Avatar}
\begin{figure}
	\centering  \includegraphics[width=0.8\linewidth]{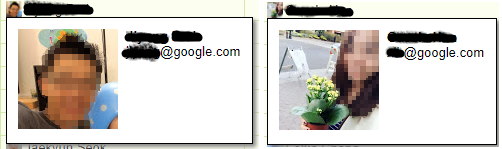}
	\caption{Gender resolution using Gerrit profile pictures: 1) male (left), 2) female (right)}
	\label{fig:gerrit-pp}	
    \vspace{-12pt}
\end{figure}

 Gerrit allows a user to include his/her picture in his/her profile. In the first step, we look into the Gerrit avatar of a user to determine his/her gender. Figure~\ref{fig:gerrit-pp} shows examples of two Gerrit avatars. The avatar on the left indicates a male contributor and the avatar on the right indicates a female. However, some users' Gerrit avatars were either empty or images that do not reveal their genders.

\subsubsection{Resolution using Google plus} 
\begin{figure}
	\centering  \includegraphics[width=0.5\linewidth]{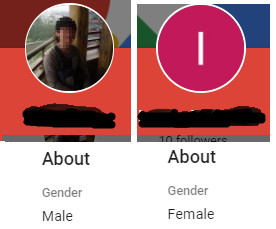}
	\caption{Gender resolution using Google plus profile: 1) male (left), 2) female (right).}
	\label{fig:google-plus}	
    \vspace{-8pt}
\end{figure}

We searched the Google plus social network using an user's email address. Based on Google plus search policy, if an user has associated his/her email address with a profile, a search based on that email address returns only that particular profile. Since the gender information of an user on Google plus is public for the majority of the users, a positive match based on an email search potentially could help finding the user's gender (Figure~\ref{fig:google-plus}). 

\subsubsection{Resolution using LinkedIn Profile}  
\begin{figure}
	\centering  \includegraphics[width=0.8\linewidth]{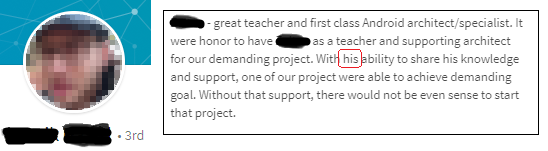}
	\caption{Gender resolution: 1) using LinkedIn profile picture (left), 2) using received recommendations (right).}
	\label{fig:linkedin}	
    \vspace{-12pt}
\end{figure}

For these users, we searched LinkedIn, a professional social network, with his/her full name and company information. For example, if a user's name is `\textit{Kai Chen}' and his/her email address is `\textit{kai.chen@intel.com}`, we searched using the term `\textit{Kai Chen + Intel}'. If we found a positive match, we inspected the profile picture to determine his/her gender. However, if a user's profile picture was invisible to us, we looked into the recommendations that he/she has received. Any gender specific pronouns (i.e., `he', `she', `his', or `her') in the recommendations revealed the user's gender. Figure~\ref{fig:linkedin} shows examples of gender resolutions using the LinkedIn.

\subsubsection{Resolution using Facebook}  We used the same search term as used on the LinkedIn (\textit{`full\_name + company\_name'}) on Facebook. If a positive match was found, we inspected the profile pictures as well as gender specific pronoun in the phrase (`To see what he/she shares..') to determine a user's gender.  

\subsubsection{Resolution using Google Search} If all the first four steps failed ($ <3\% $ overall), we searched on Google using \textit{`full\_name + company\_name'} to identify the profiles of a user on various other platforms (e.g., blog, presentation, video, Twitter, Github, and forums). If information obtained from those platforms suggests a positive match, we inspected pictures or referring pronouns on those platforms to identify the gender of a user.

\subsection{Sentiment Analysis}
Since sentiment analysis tools built for other domains do not work well on a SE dataset~\cite{ahmed2017senticr}, researchers have recently proposed several custom tools for the SE domain, such as SentiCR~\cite{ahmed2017senticr}, Senti4SD~\cite{calefato2017sentiment}, SentiStrength-SE~\cite{islam2017leveraging}, and SentiSE~\cite{sentise-tool}. Among the four tools, we use SentiSE, since i) it has the largest training dataset of 13K SE interactions among the four tools; ii) SentiSE's training dataset contains 2,800 code review comments, which are focused in this study; and iii) SentiSE boosts the highest accuracy (86.9\%), the highest weighted kappa~\cite{cohen1968weighted} (0.788), and the highest F-measure for all three classes i.e., positive (86.9\%), neutral (89.0\%), and negative (82.1\%). SentiSE~\footnote{A publication detailing the design and evaluation of SentiSE is currently under review.} 
is open source and publicly available at: 

\url{https://github.com/amiangshu/SentiSE}. 
\vspace{4pt}

We  use SentiWordNet 3.0~\cite{senti-word-net} to compile a list of words expressing sentiments. SentiWordNet is a collection of \textit{synsets}, where each synset contains one or multiple words with the same meaning. SentiWordNet identifies each synset with a unique ID, a positive score, a negative score and a definition. SentiWordNet contains 14,021 words that can potentially express sentiments (i.e. have nonzero positive/negative scores).
The objective of this research (i.e. RQ2) also requires grouping the sentiment words into categories. Among the existing categorizations, Arnold proposed the first scheme with eleven fundamental emotions~\cite{arnold-categorization}. Later, Plutchik~\cite{plutchik-categorization} and Parrot~\cite{parrott2001emotions} published reduced schemes with eight groups and six groups respectively.
Combining the three aforementioned classifications, SentiSense~\cite{senti-sense} proposes 14 emotional categories and classifies 5,496 commonly used sentiment words from the SentiWordNet.
Table ~\ref{table:sentimetExamples} shows example words from the 14 SentiSense categories.

We leverage the list of emoticons and swear words / expletives  compiled in SentiSE~\cite{sentise-tool}. Our list of swear words / expletives  include 84 commonly used ones from the english language and list of emoticons include  total 107 emoticons with 46 indicating positive sentiments and the remaining 61 indicating negatives.

\begin{table}
	\caption{Example words from the 14 SentiSense categories}
	\label{table:sentimetExamples}
	\resizebox{\linewidth}{!}{
\begin{tabular}{|l|l|}
\hline
\graycell{Category} & 
\graycell{Example words}\\
\hline
Ambiguous & ironic, rare, strong, thrilling, uncommon \\ 
\hline 
Anger & danger, disturbing, intolerable, troubling, worrisome \\ 
\hline 
Anticipation & ambition, aspiration, fair, reasonable, sufficient  \\ 
\hline 
Calmness & innocent, patient, placid, resolute, smooth  \\ 

\hline 
Despair & despair, discourage, hopeless, pessimism, unsupportive \\ 

\hline 
Disgust & confusion, disappointing, disgraceful, insane, mistake \\ 

\hline 
Fear & disastrous, horrible, rage, scary, vicious \\ 

\hline 
Hate & abysmally, crappy, rotten, shitty, worthless \\ 

\hline 
Hope & credible, encouraging, exalted, optimistic, sublime \\ 

\hline 
Joy & cheer, congratulate, fortuitous, jubilant, satisfactory \\ 

\hline 
Like & advantage, awesome, excellent, great, magnificent \\ 

\hline 
Love & adorable, kudos, lovely, marvelous, wonderful \\ 

\hline 
Sadness & alas, cry, doomed, regret, unlucky \\ 

\hline 
Surprise & amazed, astonishing, misleading, surprising, wonder\\
\hline
\end{tabular} 

}
\end{table}
   \vspace{-12pt}

\section{Results}
\label{sec-results}
In this section, we present the results of our analyses to answer the five research questions introduced in Section~\ref{sec-research-question}.  Table~\ref{table:stats} shows the results of Chi-Square tests for the five research questions to find statistical significance of the differences between male and female developers.
\begin{table*}
	\caption{Results of the statistical tests for our five research questions}
	\label{table:stats}
	\resizebox{\textwidth}{!}{
\begin{tabular}{|l|rr|rr|rr|rr|rr|}
\hline 

\multirow{2}{*}{Project} & \multicolumn{2}{c|}{ RQ1: Sentiment} & \multicolumn{2}{C{2.8cm}|}{ RQ2: Cross-gender sentiment } & \multicolumn{2}{C{2.8cm}|}{RQ3: Types of sentiment words}  & \multicolumn{2}{C{2.8cm}|}{RQ4: Emoticons} & \multicolumn{2}{C{2.8cm}|}{RQ5: Expletives / Swear words }     \\ 
\hhline{~----------} 
 & $\chi^2$ & $p$  &  $\chi^2$ & $p$ & $\chi^2$ & $p$ & $\chi^2$ & $p$ &  $\chi^2$ & $p$ \\ 
\hline 
Android & 27.98 & $<0.01$*  & 34.18& $<0.01$*&  249.62 & $<0.01$* & 38.37 & $<0.01$* & 5.70 & $0.02$*     \\ 
\hline 
Chromium OS & 182.06 & $<0.01$* & 211.03 & $<0.01$* &  361.35 & $<0.01$* & 0.005 & $0.94$ & 21.35 & $<0.01$*   \\ 
\hline 
Couchbase & 20.74 & $<0.01$* & 20.24& $<0.01$* &  347.02 & $<0.01$* & 71.79 & $<0.01$* & 4.07 & $0.04$*   \\ 
\hline 
OmapZoom & 2.27 & $0.32$ & 14.15& $0.02$*  &     1467.4 & $<0.01$* & 0.08 & $0.78$ & 0.27 & $0.60$  \\ 
\hline 
oVirt & 188.81 & $<0.01$* & 165.71& $<0.01$*   & 225.87 & $<0.01$* & 2810.2 & $<0.01$* & 3.52 & $0.05$*  \\ 
\hline 
Qt & 98.98 & $<0.01$* & 96.18& $<0.01$*  &  418.52 & $<0.01$* & 29.07 & $<0.01$* & 4.12 & $0.04$*  \\ 
\hline 
\multicolumn{10}{l}{*-statistically significant}
\end{tabular} 

}
\end{table*}

\subsection{RQ1: Gender vs. Sentiment}

\begin{figure}
	\centering  \includegraphics[width=\linewidth]{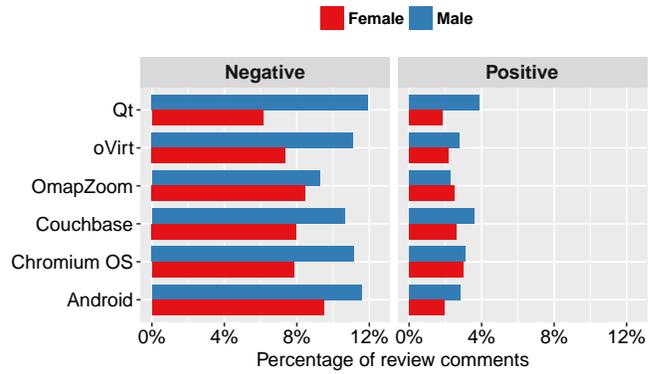}
	\caption{Distribution of sentiments: Male vs. Female}
	\label{fig:sentiment-distribution}	
    \vspace{-8pt}
\end{figure}

Figure~\ref{fig:sentiment-distribution}, shows the ratios of negative and positive review comments authored by male and female  developers from the six projects. More than 85\% code review comments in those projects are neutrals (i.e., do not express any sentiments). Among the review comments expressing sentiments, the ratios of negative sentiments are higher than positives for all six projects. For example, around 12\% reviews authored by the male developers from Android expressed negative sentiments compared to less than  3\% positives. Since the primary goal of code reviews is to identify mistakes in code, a higher ratio of negative sentiments than positives may not be surprising. 

Five out of the six projects (i.e., except OmapZoom) indicate significantly (Table~\ref{table:stats}: RQ1) higher likelihood  of review comments  authored by males expressing either negative or positive sentiments than those authored by females. 
For example, in Qt, around 4\% of the review comments authored by males were positives compared to only 2\% from females. Similarly, $\approx$12\% reviews from males were negatives compared to $\approx$6\% from females.
We also found more than 90\% review comments authored by female developers as neutrals.
    \vspace{-6pt}
\begin{boxedtext}
\textbf{Finding 1:} \emph{Male developers were significantly more likely to author review comments expressing positive / negative  sentiments than females.}
\end{boxedtext}
 
\begin{figure}
	\centering  \includegraphics[width=\linewidth]{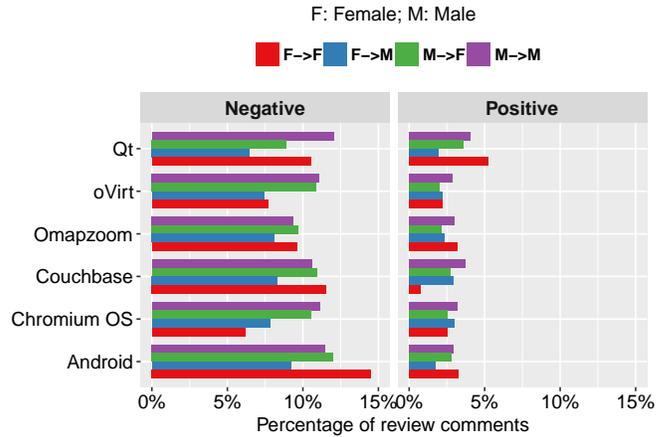}
	\caption{Distribution of sentiments: Same gender vs. cross-gender }
	\label{fig:group-sentiment-distribution}	
    \vspace{-8pt}
\end{figure}

\subsection{RQ2: Same Gender Vs. Cross-Gender Interaction}
\label{sec-rq2}
We computed the distributions (Figure~\ref{fig:group-sentiment-distribution}) of negative and positive review comments during male $\rightarrow$ male, male $\rightarrow$ female, female $\rightarrow$ male, and female $\rightarrow$ female interactions. Our results suggest significant differences among those distributions for all the six projects (Table~\ref{table:stats}: RQ2). Four out of the six projects (ie., except Couchbase and Chromium OS) show females more frequently expressing  both positive and negative sentiments during  their interactions with another female than during than interactions with a male. In Couchbase, females were more negative but less positive when communicating with another female. However, Chromium OS shows an exception with females expressing less sentiments  to other females than to males. 

For males, the results are mixed.  In Qt, Chromium, and oVirt males expressed more sentiments, both positives and negatives, during  their interactions with another male than during their interactions with a female. But  male developers seem to be harsh to their female colleagues in Android, Couchbase, and Omapzoom,  as males not only wrote negative reviews more frequently but also wrote positive reviews less frequently during their interactions with a female than during their interactions with another male. 
    \vspace{-6pt}
 \begin{boxedtext}
\textbf{Finding 2:} \emph{ In five of the six projects, female developers were more likely to express sentiments  to another female than to a male. However, in three out of the six projects, males were harsher to females by not only providing more negative reviews but also providing less positive encouragements.}
\end{boxedtext}

\begin{figure}
	\centering  \includegraphics[width=\linewidth]{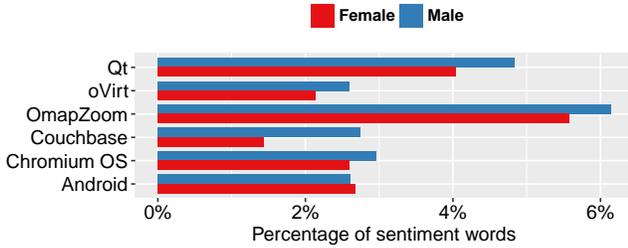}
	\caption{Frequency of sentiment words:  Male vs. Female}
	\label{fig:senti-words}	
    \vspace{-8pt}
\end{figure}

\begin{figure*}
	\centering  \includegraphics[width=\linewidth]{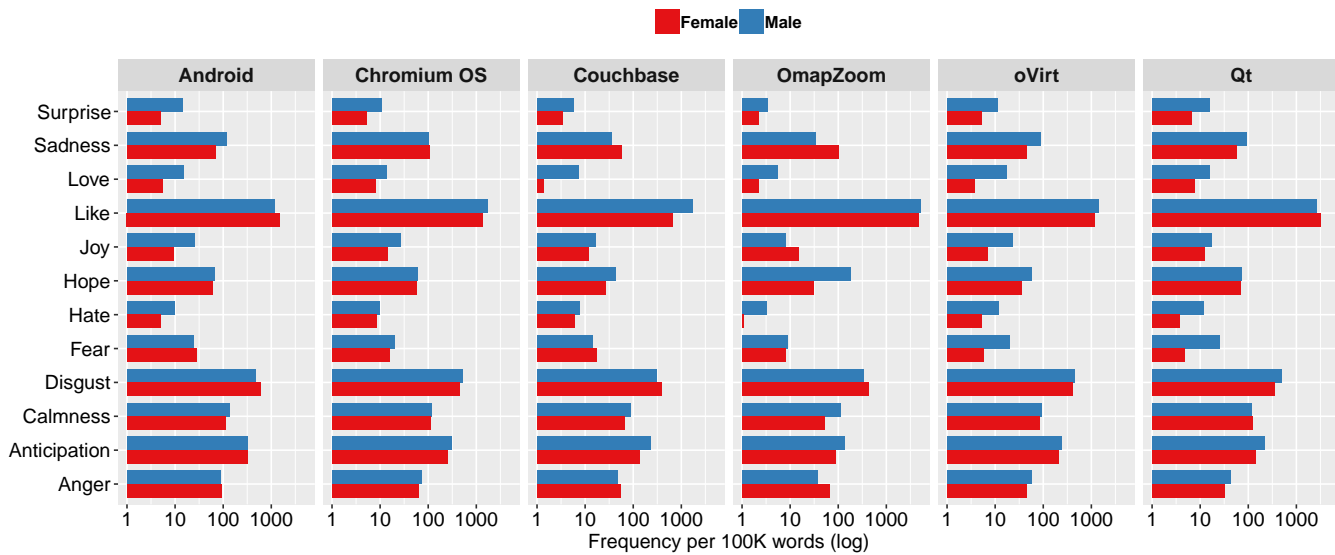}
	\caption{Distribution of sentiment word categories from SentiSense: Male vs. Female}
	\label{fig:word-distribution}	
    \vspace{-8pt}
\end{figure*}

\subsection{RQ3: Gender Vs. Sentiment Word}
\label{sec-rq3}
We computed the frequencies of each word belonging to the \textit{SentiSense}~\cite{senti-sense} among  the reviews authored by both male and female developers.
Among the 14 categories from \textit{SentiSense}, words belonging to the `ambiguous' category assume sentiment orientations based on its context. Moreover, words belonging to the `despair' category were rare or even non-occurring among the review comments in our dataset. Therefore, we excluded these two categories from our analysis. 

In five of  the six projects (except Android) male developers used sentiment words more  frequently than females (Figure~\ref{fig:senti-words}).
Figure~\ref{fig:word-distribution} shows the occurrences of the 12 SentiSense categories per 100K words authored by male and female developers  from the six projects. Since words from some categories were several times more frequent than the other categories, we plot those charts on a log scale. For example, males from Android used words from the `like' category 1,510 times per 100k words, but the corresponding number for the words from the `fear' category was only 23. We also noticed significant differences between males and females  in using  \textit{SentiSense words} (Table~\ref{table:stats}: RQ3).

In three of the six projects (i.e., except Android, Couchbase, and OmapZoom), males were more likely to use words belonging to strong sentiment categories such as `Surprise', 'Sadness', 'Love', `Joy', `Hate', `Disgust','Fear', and `Anger'. 
The frequencies for the words belonging to mild sentiment categories such as `Like', `Anticipation', `Calmness' and `Hope'  were very similar between males and females. Females from Android, Couchbase and Omapzoom were more frequently expressing `Anger' and `Disgust' possibly due to the negative attitudes from their male colleagues.

    \vspace{-6pt}
\begin{boxedtext}
\textbf{Finding 3:} \emph{ Male developers were significantly more likely to use sentiment words than females. Moreover, during sentiment expressions, females were less likely to use words expressing strong sentiments than males.}
\end{boxedtext}

\begin{figure}
	\centering  \includegraphics[width=\linewidth]{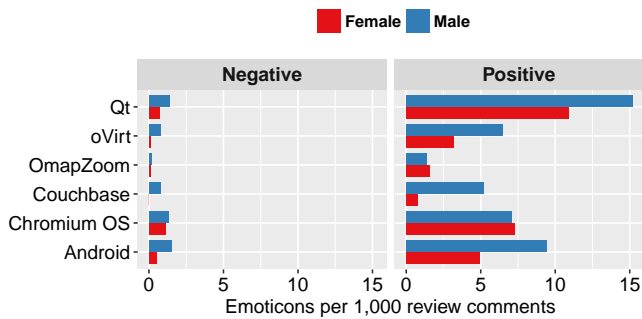}
	\caption{Distribution of emoticons:  Male vs. Female}
	\label{fig:emo-distribution}	
    \vspace{-8pt}
\end{figure}

\subsection{RQ4: Gender Vs. Emoticon}
Figure~\ref{fig:emo-distribution} shows the distributions of positive / negative emoticons per 1000 review comments authored by male and  female developers. In general, developers use positive emoticons far more frequently than negatives. The  two `smiley face' emoticons (i.e., `:)' and `:-)') accounted for more than 75\% of the emoticon usage for each project. Among the negative emoticons, sad face `:(' and toungue out `:p' were the most frequent ones. We also noticed emoticons more frequent among comments from males than from females (except positive emoticons in Chromium OS). These differences between male and female developers in using emoticons are statistically significant for four out of the six projects (except Chromium OS and OmapZoom) as shown in Table~\ref{table:stats}: RQ4. However, we did not notice any significant affinity towards a particular emoticon based on gender.
    \vspace{-6pt}

\begin{boxedtext}
\textbf{Finding 4:} \emph{ Contrary to prior results, female developers were less likely to use emoticons than males during code reviews. }
\end{boxedtext}
 
\subsection{RQ5: Gender Vs. Expletive / Swear Word}
\label{sec-rq5}
\begin{figure}
	\centering  \includegraphics[width=\linewidth]{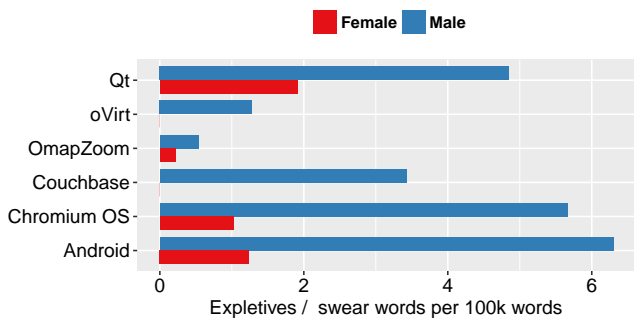}
	\caption{Distribution of expletives / swear words: Male vs. Female}
	\label{fig:slang-distribution}	
     \vspace{-8pt}
    \end{figure}

Figure~\ref{fig:slang-distribution} shows the distributions of expletives / swear words  per 100k words authored by male and  female developers. In general, the frequencies of expletives / swear words were very low, which may not be surprising since our dataset includes six of the top OSS projects. We also noticed male developers more frequently using expletives / swear words than females. These differences are also statistically significant for five out of the six projects (i.e., except OmapZoom) as shown in (Table~\ref{table:stats}: RQ5).
We also noticed `crap', `damn', and `screw' as the most common expletives used by both males and females.  However, some of the highly offensive expletives such as `bitch', `bastard', `fuck', `jerk' were used only by males.

    \vspace{-16pt}
\begin{boxedtext}
\textbf{Finding 5:} \emph{Female developers were significantly less likely to use expletives than males in five out of six projects. Even when female used expletives, they avoided certain highly offensive ones that males often used.}
\end{boxedtext}

\section{Discussion and Implications}
\label{sec-discussion}
In this section, we compare our results with prior results from other domains and discuss possible reasons and implications for our findings. 
\subsection{Verbal Abuses Towards Females}
One of the primary goal of this study is to determine whether females are more frequent recipients of negative opinions than males. The results are mixed, since three out of the six projects indicate possible discrimination with males not only writing negative reviews more frequently to females but also withholding positive encouragements (Section~\ref{sec-rq2}). Unsurprisingly, females from those projects were more frequently using words expressing `Anger' or `Disgust' than the males belonging to those projects (Section~\ref{sec-rq3}) in their reviews. However, it is encouraging to see this trend as not a norm across all the six projects.

On the other hand, the results of RQ5 (Section~\ref{sec-rq5}), which suggest more frequent expletives from the males than from the females, with some of expletives being highly offensive to females, are concerning. Some software projects, especially OSS projects have been guilty of fostering a `toxic culture' for women, where inflammatory talk and aggressive posturing is acceptable within the norm of the community~\cite{nafus2006gender}. Some words often used in the mailinglists are insulting to women. Recently similar toxic culture has been also reported in Apple~\cite{apple}, Tesla~\cite{tesla} and UploadVR~\cite{uploadvr}, where crude and sexist jokes often made female employees uncomfortable. 
A similar scenario seems to be present across all the six projects in our study.

\subsection{Comparisons with Other Domains}
While some of our results support prior findings in other areas, several of our results are contradicting. 
First, while women have been termed as more emotional than men in most of the prior studies~\cite{montero2014investigating,thelwall2010data,zammuner2000men},  we find a completely opposite picture in the SE context.
Second, Thelwal et al.~\cite{thelwall2010data}, found male-female interactions to be slightly flirtatious, while female-female interactions to be transparently supportive. On the contrary, our findings indicate that females remain neutral while interacting with males, but males render more negative reviews to their female peers. Although females writing more positive comments to another female may hint support, females also wrote more negative reviews to another female.
Third, prior studies found a tendency for women to report feeling stronger and longer emotions and to express them more clearly~\cite{fischer2000relation}. However, five out of our six projects indicate a contradicting picture with males using stronger emotions more frequently than females.
Fourth, in terms of emoticon usage, Wolf found women more frequently using emoticons than men. While women used humors more frequently, emoticons to express teasing or sarcasm dominated in males' newsgroups~\cite{wolf2000emotional}. However, our findings not only found less emoticons from females but also did not find any significant differences based on gender.
Finally, although recent studies suggest~\cite{stapleton2003gender, ott} females breaking traditional stereotype in the usage of expletives or swear words, we did not observe a similar trend in the SE context.

\subsection{Non-expressiveness of Female Developers}

One intriguing question gets raised by our results is, \textit{why female developers do not express their sentiments in the SE context as they often do in social media}. While we do not have a definite answer, several factors, as discussed following may be contributing.
Inclusive communities with more women can be more comfortable for women to ask questions or interact with each other. However women constitute less than 10\% OSS developers~\cite{FOSS}. 
A recent survey on Github developers also report profound gender imbalance with 95\%  males compared to only 3\% females~\cite{github-survey}. 
As a result, women may be misinterpreted in software engineering while expressing their feelings or sentiments. A woman who speaks up may be judged as aggressive in OSS projects~\cite{FOSS} and therefore, in most of the cases, women keep themselves silent and do not speak up or show emotions. 
In another survey~\cite{Harvard}, female executives reported that their comments during heated discussions are misinterpreted and are perceived as emotional. Therefore, women were advised to back up their feelings with logic, specificity and facts.

The  uneasiness or self-doubt and less confidence in female developers may also arise from the  lack of female mentors or seniors in software engineering. For example, women held only 11\%  of top executive positions at Silicon Valley companies~\cite{female-exec}. Moreover, women were three-and-half times more likely to holding junior positions, in spite of being equally capable as their male peers~\cite{cnet}. 
However, the presence of female role models, as recently observed in Malaysia~\cite{Ulf2009}, may  make the SE industry less associated with masculine characteristics and more appropriate for women. Sheryl Sandberg describes this issue as  ``\textit{There aren't more women in tech because there aren't more women in tech.}~\cite{lansing}." 

In a recent survey by Elephant in the Valley~\cite{elephant}, 87\% of the women working in the Silicon Valley reported demeaning comments from their male colleagues. More alarmingly, 60\% of the women reported sexual harassments, which is 1.5 times higher than the percentage of women reporting the same in other areas (41\%)~\cite{das2009sexual}.
Moreover, women are often questioned on their moral standing and face social consequences for their linguistic behavior. Using aggressive or harsh or forceful words is treated as infraction of their cultural and social stereotypes. Using expletives breaches behavioral compliance in some communities and society imposes more obligations on women in preserving social values and norms than men~\cite{stapleton2003gender}.
Due to these factors,  female developers who do not feel comfortable at their work environments, may subdue their natural instincts~\cite{zammuner2000men} to express emotions.

\subsection{Implications}

In recent years the US government as well as IT organizations have taken numerous initiatives to increase the enrollments of females in computing education~\cite{csforall, girswhocode, nasagirls}, which helped to improve the percentage female CIS graduates to 18\% in 2016~\cite{taulbee2016} from 12\% in 2006. 
However, retaining women in computing professions received very little attention~\cite{panteli1999status}.
A recent report suggests that 45\%  women who chose computing careers leave the field within ten years and that quit rate is more than twice as high for women than it is for men~\cite{ashcraft2016women}. 
The primary reasons behind the attrition can be negative workplace experiences, lack of access to creative technical roles, and dissatisfaction with career prospects~\cite{ashcraft2016women}. 
As a result, although the percentage of female graduates continued to increase during the last decade, the percentage of women holding computing jobs are declining~\cite{anna2016}. According to a report from the US Department of Commerce, women accounted for 30\% of the computing jobs in 2000 and that number fell to 27\% in 2009~\cite{beede2011women}. Another report by the ``Girls Who Code" initiative estimated the percentage of women in computing jobs steady at 24\% from 2011 to 2016 despite the growing ratio of female graduates and that number will likely reduce to 22\% by year  2025~\cite{girswhocode}. 
Being more frequent recipients of negative reviews than males (Section~\ref{sec-rq2}) or encountering words that are demeaning to females (Section~\ref{sec-rq5}) may be a factor behind the negative workplace experiences encountered by the female developers and may be contributing to the diminishing number of females in the SE domain.

A recent study~\cite{Cheryan2017} investigated the gender biasness in six of the largest natural science and engineering fields. The results found biological sciences, chemistry, mathematics, and statistics as more gender-balanced compared to computer science, engineering, and physics. According to Cheryan \textit{et} al. the three dominant factors shaping preferences for one subfield over another  are:  i) insufficient early experience,  ii) perceptions of a masculine culture, and  iii) gender gaps in self-efficacy.  Some computing organizations foster `geek culture'~\cite{margolis2003unlocking}, which promotes certain stereotypes that do not fit well with most women. Therefore, women often feel as outsiders within those organizations. The characteristics and culture in communication among male and female counterparts in code reviews can represent the true nature of their inter relationship and the masculine culture in SE fields. So, this study will open up opportunity of research on gender bias in SE field from another point of view. Our study reveals the true nature of the behavioral differences between men and women in the SE field and further investigations can help overcoming diminishing number of females in this domain~\cite{girswhocode}.

\section{Threats}
\label{sec-threats}
In the following subsections, we address three common types of threats to any empirical study.

\subsection{Internal Validity:} 
The primary threat to internal validity in this study is \textit{project selection}.
We included six publicly accessible OSS projects that practice tool-based code reviews supported by the same tool (i.e., Gerrit). 
Though, it is possible that projects supported by other code review tools (e.g., ReviewBoard, Github pull-based reviews, and Phabricator) could have behaved differently, we think this threat is
minimal for four reasons:  1) all code review tools support the same
basic purpose, i.e. detecting defects and improving the code, 
2) the basic workflow (i.e. authors posting code, reviewers commenting
about code snippets, and code requiring approval from reviewer before integration) of most of the code review tools are similar,  3) we did not use any Gerrit-specific feature/attributes in this study,  and 4) sentiments expressed in review comments may not depend on any feature that is exclusive to Gerrit only. Therefore, we believe the project selection threat is minimal.

\subsection{Construct Validity:} 
The primary threat to construct validity is related to our gender resolution methodology based on the genderComputer tool, which has been used in several prior SE studies~\cite{vasilescu2015gender,terrell2017gender}. Since we manually  validated the tool's classification for the 937 non-males, we believe that no male was misclassified as a female in our dataset. However, as we accepted genderComputer's classifications for the 2,633 males, we are unable to make a similar claim that no female was misclassified as a male. To estimate, possible classification errors, we manually validated  the genders of 200 randomly selected developers from the 2,633 males, using a similar methodology that we used for the non-males. Since our validation found only one female (0.5\%) among these 200 developers, we do not think  enough females were misclassified as males in our dataset to alter the results of our research questions.

We use SentiSE, a state-of-the-art sentiment analysis tool which is highly accurate ($\approx$87\%). However, we cannot confidently claim that the misclassification ($\approx$ 13\%) of SentiSE has not altered the results of our study. However, this would have been true only if  there is a systematic relationship between the comments that SentiSE
incorrectly classifies and the gender of the author (if, for example, SentiSE incorrectly classifies
comments written by females more frequently than comments written by males). We have no reason to believe that such a relationship exists, but have no empirical evidence.
 
\subsection{External Validity}
Although we analyzed a large number of code review requests from six popular and matured OSS projects, we cannot definitively establish that our sample is representative of the entire OSS population. 
Because OSS projects vary on characteristics like product, participant type, community structure, and governance, we cannot draw general conclusions about all OSS projects from this single study. To build reliable empirical knowledge, we need
family of experiments~\cite{basili1999building} that include OSS projects of all types.

\section{Conclusion}
\label{sec-conclusion}
In this study, we explored the differences between men and women in using sentiment words, emoticons, and expletive during code reviews. 
This work is motivated by earlier findings where authors investigated existing gender bias in tech fields and differences of sentimental expression of men and women in other contexts such as social media. 
However, no prior research has explored the differences in opinions between men and women in the SE domain, and broadly in tech fields. 
Our results suggest that the likelihood of using sentiment words, emoticons, and expletives during code reviews vary based on the  gender of a developer as  women are less likely to use sentiment words / emoticons / expletives than men. We also investigated same-gender and cross-gender interactions and found female developers less frequently writing negative comments to  their male counterparts. Yet, male developers from three out of the six projects  were not only critical of their female counterparts but also withheld positive encouragements. Unsurprisingly, females from those projects were expressing `Anger' and `Disgust' more frequently than males. Our results also found males more frequently using expletives or words that are demeaning to females across all the six projects, supporting a `toxic culture' as suggested by a prior study~\cite{nafus2006gender}.   The results of this research can be used to analyze the ongoing gender issues in tech fields from different perspective. 
The diminishing trend of gender gap in other fields including automotive industry, biological sciences and mathematics~\cite{Cheryan2017} stems from the pursue of investigating the gender issues from the very beginning of education level to the professional environment.
Therefore, our results will be interesting for the researchers to investigate the current socio-cultural norms against women in tech fields and for the professionals to create a workplace where every woman can feel confident, supported and safe to pursue their dreams. 
In a follow-up study, we plan to further investigate the outcomes of this study qualitatively and determine possible factors  behind our results. We hope our results can motivate further research  both to retain and to encourage more women in computing professions.

\bibliographystyle{IEEEtranS}

\bibliography{references}

\end{document}